\documentclass[useAMS,usenatbib,usegraphicx]{mn2e}

\usepackage{graphicx}
\usepackage{txfonts}

\newcommand{\apj}{ApJ}

\def\wb{\mbox{$B_{435}$}}

\def\wi{\mbox{$i_{775}$}}
\def\wz{\mbox{$z_{850}$}}
\def\lesssim{\mathrel{\hbox{\rlap{\hbox{\lower4pt\hbox{$\sim$}}}\hbox{$<$}}}}
\def\gtrsim{\mathrel{\hbox{\rlap{\hbox{\lower4pt\hbox{$\sim$}}}\hbox{$>$}}}}
\def\see{\mbox{$^{\prime\prime}$}}

\title[Ionising radiation evolution]{Contamination on Lyman continuum 
emission at z $\gtrsim$ 3:\\ implication on the ionising radiation evolution}

\author[Vanzella et al.]{E. Vanzella$^{1}$\thanks{E-mail: vanzella@oats.inaf.it -- 
Based on observations made at the European Southern
Observatory, Paranal, Chile (ESO programme 170.A-0788 {\it The Great
Observatories Origins Deep Survey: ESO Public Observations of the
SIRTF Legacy/HST Treasury/Chandra Deep Field South.})},
B. Siana$^{2}$,  
S. Cristiani$^{1}$ and
M. Nonino$^{1}$
\\
$^{1}$ INAF - Osservatorio Astronomico di Trieste, Via G.B. Tiepolo 11, 40131 Trieste, Italy.\\
$^{2}$ California Institute of Technology, MS 105-24, Pasadena, CA 91125, USA.
}

\begin{document}
\pagerange{\pageref{firstpage}--\pageref{lastpage}} \pubyear{2009}
\maketitle
\label{firstpage}

\begin{abstract}
We investigate the possibility of contamination by lower-redshift interlopers
in the measure of the ionising radiation escaping from high redshift galaxies.
Taking advantage of the new ultra-deep VLT/VIMOS U-band number counts in the GOODS-South field,
we calculate the expected probability of contamination by low redshift interlopers as a 
function of the U magnitude and the image spatial resolution (PSF).
Assuming that ground-based imaging or spectroscopy can not resolve objects lying within
a 0.5\see~radius of each other, then each z$\gtrsim$3 galaxy has a 2.1 and 3.2\%~chance
of foreground contamination, adopting surface density U-band number counts down to 27.5 and 28.5,
respectively. Those probabilities increase to 8.5 and 12.6\%, assuming 1.0\see~radius.
If applied to the estimates reported in the literature at redshift $\sim$ 3 for which a Lyman continuum
has been observed $directly$, the probability that at least 1/3 of them are affected by foreground contamination
is larger than 50\%. 
From a Monte-Carlo simulation we estimate the median integrated contribution of foreground sources to 
the Lyman continuum flux ($f900$). 
Current estimations from stacked data are $>$2$\sigma$ of the median integrated pollution by foreground sources.
If the correction to the observed $f900$ flux is applied, the relative escape fraction 
decreases by a factor of $\sim$ 1.3 and 2, depending on the cases reported in literature.
The spatial cross-correlation between the U-band ultra-deep catalog
and a sample of galaxies at z$\gtrsim$3.4 in the GOODS-South field, produces a number of
U-band detected systems fully consistent with the expected superposition statistics. Indeed, each of them
shows the presence of at least one offset contaminant in the ACS images.
An exemplary case of a foreground contamination in the Hubble Ultra Deep Field at redshift 3.797
by a foreground blue compact source (U=28.63$\pm$0.2) is reported; if observed with
a low resolution image (seeing larger than 0.5\see) the polluting source would mimic
an observed $(f1500/f900)_{OBS}\sim$ 38, erroneously ascribed to the source at higher redshift.
\end{abstract}

\begin{keywords}
Galaxies: evolution -- Galaxies: high-redshift -- Individual source: GDS~J033236.83-274558.0
\end{keywords}

\section{Introduction}
\label{intro}
Ionising radiation from star-forming galaxies is a plausible primary source of cosmic reionisation.
The amount, the evolution with redshift 
and its contribution to the cosmic reionisation 
is still a matter of debate (e.g., Steidel et al. (2001), Giallongo et al. (2002), 
Shapley et al. (2006) [S06, hereafter], Siana et al. (2007, 2010), Faucher-Gigu\`ere et al. (2008), 
Iwata et al. (2009) [I09, hereafter]).

Malkan et al. (2003) and Siana et al. (2007, 2010) have stacked tens of deep ultraviolet 
imaged galaxies at redshift $\sim$ 1 and no detection has been reported. Similarly,
Cowie et al. (2009) have combined $\sim$ 600 galaxies observed with GALEX at the same redshift 
and also in this case the result was a non detection. At higher redshift (z$>$3) an escape
of Lyman continuum photons (LyC hereafter) seems to be detected in $\sim$ 10\% of the 
galaxies (e.g. I09 and S06), suggesting a possible evolution of the escape fraction with redshift.
It is still unclear which is the typical escape fraction to be expected from starburst galaxies 
at redshift z$\gtrsim$3. While some results of numerical simulations suggest low values 
(less than 10\%, e.g. Gnedin et al. 2008), others claim escape fraction as 
high as 80\% (e.g., Yajima et al. 2009, Razoumov \& Sommer-Larsen 2009).
The two possibilities imply two completely different scenarios: in the first case 
high-redshift galaxies turn out to be inefficient in releasing ionising radiation into the 
intergalactic medium, conversely in the second case, they may play a dominant role in this redshift regime.

As redshift increases, the probability of foreground contamination by blue galaxies that mimic
the ionising radiation also increases.
Therefore it is important to study the contaminated systems and try to quantify their occurrence.
The recent ultra deep VLT/VIMOS U-band observations in the GOODS-S field (Nonino et al. 2009) 
allow to investigate this issue to unprecedented depth and to better calculate the probability of such contamination
(a comparison between the observed superpositions in the GOODS-S area and those expected is also 
performed). Very deep exposures with exquisite spatial resolution, as the Hubble Ultra Deep Field (HUDF, Beckwith et al. 2006),
make it possible to verify the results of such a calculation over limited areas of the sky.
The work is structured as follows: in  Sect. 2.1 the probability distributions of foreground superposition
are calculated and in Sect. 2.2 it is compared to the observed direct detections reported in literature. 
In Sect. 2.3 the Monte-Carlo simulations are described to derive the integrated contribution of foreground sources.
In Sect. 3 the U-band catalog is cross-matched to the redshift $\sim$ 4 sample and checked for contamination. 
Sect. 3.1 discuss the case of an exemplary $false$ LyC escape detection in the HUDF. Sect. 4 summarise the results.
The standard cosmology is adopted ($H_{0}$=70 $km~s^{-1}~Mpc^{-1}$, $\Omega_{M}$=0.3, $\Omega_{\Lambda}$=0.7).
If not specified, magnitudes are given in AB system.

\section{Lyman continuum escape radiation at z$>$3}
\label{probs_direct}
\begin{figure} \centering
  \resizebox{\hsize}{!}{\includegraphics{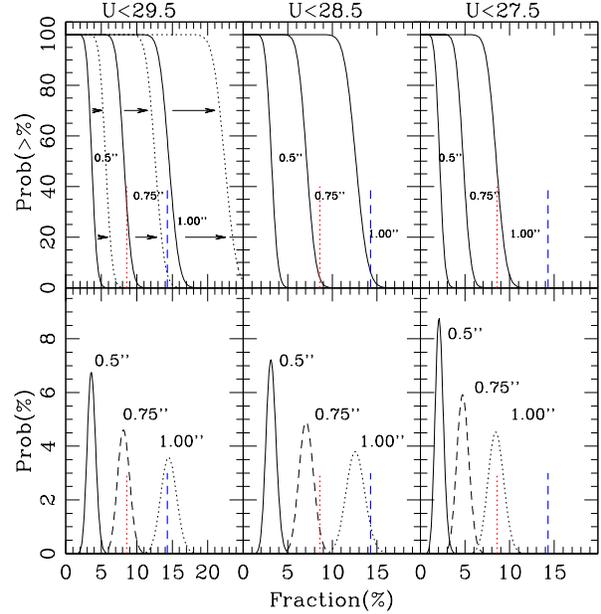}}
\caption{Probabilities (Prob, Y-axis) that a fraction (\%, X-axis) of an observed sample is contaminated by foreground sources. 
Calculations have been performed varying the single case probability, $p$, in turns
related to the U-band number counts (from left to right panels) and seeing (different curves in each panel). 
{\bf Top panels:} The cumulative probability distribution 
versus fractions (\%) of the entire sample ($N$). It is the probability that $at~least$ a fraction $K$(\%) of the $N$ 
galaxies observed (P($\geq$K) with $N>K$) is contaminated by foreground sources.
In the fainter limit case, U-mag $<$ 29.5 (left panel), calculations have been performed with the 
observed number counts (solid lines) and those corrected 
by uncompleteness in the range 28.5$<$U-mag$<$29.5 (dotted lines, see text for details). 
The arrows show how the curves change from uncorrected to corrected.
{\bf Bottom panels:} The probability distributions that a fraction $K$(\%) of the $N$ galaxies observed (P(K)) are 
 contaminated. Increasing the U-band surface density and the seeing, the probability of superposition increases.
In all panels the two vertical dotted (red) and dashed (blue) lines mark the percentages of the direct 
LyC detections by I09 (17/198) and S06 (2/14), respectively.}
\label{chance}
\end{figure}
\begin{table}
\centering 
\caption{Number counts derived from the ultradeep VLT/VIMOS catalog down to U $\sim$ 30 (1$\sigma$ limit within
1\see~aperture radius, Nonino et al. 2009). Completeness correction has been roughly applied at magnitudes fainter 
than 28.5 ($\sim$3$\sigma$ limit). The values in parenthesis have been calculated by extrapolating linearly the number
counts up to U-mag 30.5 by fitting a straight line in the range U-mag 25 -- 28 with a slope 0.2 in the [$LogN$] vs. [U-mag] plan.}
\label{U_COUNTS} 
\begin{tabular}{lll} 
\hline
\hline   
         U-mag          &  Ncounts            &  Cumul   \\
        Ulow -- Uup             & [$deg^{-2}$] & ($m<$ Uup) [$deg^{-2}$]\\
\hline
              22.0 -- 22.5     &1400$\pm$40      &1400\\
              22.5 -- 23.0     &2200$\pm$50      &3600\\
              23.0 -- 23.5     &4400$\pm$70     &8000\\
              23.5 -- 24.0     &8300$\pm$90     &16300\\
              24.0 -- 24.5     &14900$\pm$120     &31200\\
              24.5 -- 25.0     &23800$\pm$150     &55000\\
              25.0 -- 25.5     &33500$\pm$180     &88500\\
              25.5 -- 26.0     &45600$\pm$210    &134000\\
              26.0 -- 26.5     &57900$\pm$240    &191900\\
              26.5 -- 27.0     &72300$\pm$270    &264200\\
              27.0 -- 27.5     &85300$\pm$270    &349500\\
              27.5 -- 28.0     &90200$\pm$300    &439700\\
              28.0 -- 28.5     &80400$\pm$280    &520200\\
\hline
              28.5 -- 29.0     &54900$\pm$230 (177800)    &575100 (698000)\\
              29.0 -- 29.5     &23700$\pm$150 (223900)    &598800 (921900)\\
              29.5 -- 30.0     &4500$\pm$70 (281800)    &603300 (1203700)\\
              30.0 -- 30.5     &---~~(354800)        &603300 (1558500)\\
\hline
\hline
\end{tabular}
\end{table}

\subsection{The expected probability of foreground contamination}
\label{calc_probs}
Taking advantage of the ultra-deep VIMOS U-band imaging in the GOODS-S field (Nonino et al. 2009), that
reaches magnitude 29.8 (28.5) at 1$\sigma$ (3$\sigma$) within 1\see~aperture (radius), we 
determine the likelihood of foreground 
contamination of z$>$3, as these galaxies must reside in the foreground and are emitting at wavelengths 
that mimic LyC of LBGs. Similar to the discussion in Siana et al. (2007) we assume 
that the spatial distribution of the foreground (U-detected) galaxies is uniform and uncorrelated with 
those at higher redshift. 
If we adopt the surface density of objects with U(AB)$<$28.5 (520200 $deg^{- 2}$, see Table~\ref{U_COUNTS} 
and Nonino et al. (2009)) and
assume that ground-based imaging or spectroscopy can not resolve objects which lie within
a 0.5 (0.8)\see~radius of each other, than we would expect that 
each z$\gtrsim$3 galaxy has a $p$$\sim$3.2 (8.1)\%~chance of foreground contamination.
Figure~\ref{chance} summarise the probabilities that a certain fraction (\%) of the total sample $N$ is
contaminated by foreground sources, as a function of the seeing and U-band number counts.
 \footnote{The probability to observe $K$ (or $\ge K$) contaminated sources 
in a sample of $N$ ($>K$) high-z galaxies given the probability $p$ of the single case 
is: ~~~~~~~$f(K)=\left(^{N}_{K}\right)p^{K}(1-p)^{N-K};$ ~~~~~~~~$P(\ge K)= \sum_{i=k}^N f(i);$}
It is evident that increasing the U-band magnitude limit of galaxy counts and the seeing value,
the probability of foreground contamination increases. 
Depending on the seeing conditions and the U-band magnitude limit adopted, the fraction of the
population contaminated by foregorund sources ranges between 3 and 15\%, more severely
if the faint fluxes are investigated (see Figure~\ref{chance}).
Currently, the direct LyC detections reported in literature regard fractions smaller than 15\%.
It is worth noting that stacking of sources can also be affected by contaminantion, 
being a result of a sum that may include contaminated cases not detected individually (see Sect. 2.3).
In the following sections we consider the observed number counts down to U-mag 28.5 (3$\sigma$ limit), 
beyond this limit a linear extrapolation is performed (see Sect. 2.3).
In the next section the derived probabilities are compared to the observations of $direct$ LyC.

\subsection{The current LyC detections}
\label{obs_direct}
S06 reported on LyC detections: 2 out of 14 galaxies show a clear signal below 912\AA\ rest-frame in their deep spectra.
In particular SSA22a-C49 ($z$=3.115) and SSA22a-D3 ($z$=3.067) show a 
LyC flux of $f_{900}$=(11.8$\pm$1.1)$\times$$10^{-31}$ ergs~$s^{-1}$~$cm^{-2}$~$Hz^{-1}$ (mag(AB) = 26.22)
and $f_{900}$=(6.9$\pm$1.0)$\times$$10^{-31}$ ergs~$s^{-1}$~$cm^{-2}$~$Hz^{-1}$ (mag(AB) = 26.80), respectively.
Adopting the seeing value reported in S06, never worse than 1.0\see~and larger than 0.8\see, 
and the number counts in the magnitude range 26$<$U$<$27 (130200 $deg^{- 2}$), 
the probabilities that at least one galaxy (out of 14) in the S06 
study is subject to foreground contamination assuming seeing 0.8, 0.9 and 1.0\see, 
are 25, 30, 36\%  respectively. There is a 3, 5, 7\%~chance that at least two 
detections are contaminated. 

It is worth noting that in the case of spectroscopic observations, the slit
width (1.2\see in S06) and its orientation on the sky may influence the net contribution 
of a close (``sub-seeing'') contaminant, since it may be $partially$ included in the slit if there is 
a spatial offset. Therefore the $f900$ flux due to offset foreground sources is decreased. 
In principle, there is also the additional possibility to identify the 
foreground source through its spectral fatures if the emission lines are present in the wavelength
range and are intense enough to be detected.

I09 reported similar 
LyC detections in LBGs and LAEs at redshift 3.1 with dedicated narrow band imaging (NB359). 
In particular 17 (7 LBGs and 10 LAEs) sources out of 198 (73 LBGs and 125 LAEs) show a clear signal ($\gtrsim$3$\sigma$).
Looking at their Figure 4 the NB359 magnitude of the 17 cases is in the range 26-27.5,
the surface density of U-band detected galaxies in this interval is 215500 $deg^{- 2}$ (Nonino et al. (2009)).
Assuming as in I09 that a foreground object within 1.0\see~radius cannot be distinguished in the image,
each object has 5.2\% chance of contamination, corresponding to a 59(46)\% chance that at least 
(10/11) sources out of 198 are contaminated by foregrounds. If we adopt a U magnitude range 26.5-27.5
as in I09, the U surface density becomes 157600 $deg^{- 2}$, corresponding to a 49\% of 
probability that at least 8 objects out of 198 are contaminated.
Furthermore, I09 note that shapes and positions of the emitting regions in the sub-sample of
LBGs is different from those in the R-band, reporting an average offset of 0.97\see.
Since spatial offsets are detected, I09 used adaptive apertures (centered in the $R$ image) 
to measure the bulk of the fluxes in the $R$ and NB359 filters and derive the colour (i.e. the observed $f1500/f900$ ratios).
The aperture diameters adopted vary from 2.0\see~to 4.0\see, therefore, apart from the ``intrinsic'' confusion due 
to the seeing, the relatively large apertures may include even more contaminants.
Considering the sub-sample of 73 LBGs for which an offset has been measured, and adopting the 
median case of 1.0\see~radius and surface number density of galaxies 
with 26.5 $<$ U $<$ 27.5 of 157600 $deg^{- 2}$, it turns out that there is 53 (2) \%~chance that at 
least 3 (7) estimated colours are contaminated. With 1.5\see~radius the chance
is 96 (44)\%. 
I09 suggest that the spatial offset may be related to the intrinsic nature and/or the geometry
of the escape radiation process, mantaining however open the possibility of a superposition for part of
the sample.

Finally, it is worth noting that the two LyC detected galaxies by S06 (SSA22a-D3 and SSA22a-C49) have 
been observed also by I09. In the case of SSA22a-D3, I09 notice that the object is not visible in their NB359 image, 
even though the flux density limit is well below (two times) the limit reported by S06.
The other source, SSA22a-C49, shows a NB359 signal (2.95$\sigma$ level) spatially shifted with respect to the $R$ band. 
From the ACS/F814 imaging (shown in I09, Figure 2, top right panel) it is apparent a 
distinct substructure within 1.0\see~from the brighter LBG, spatially consistent with the emission 
in the NB359 image. It would be important to obtain the redshift of such companion.

All these considerations suggest that the situation is still unclear and even though the pure foreground 
contamination may not entirely explain the direct LyC detections at high
redshift, its effect is not negligible.

\subsection{The integrated contribution of foreground sources}
\label{Lya_integ}
\begin{figure} \centering
  \resizebox{\hsize}{!}{\includegraphics{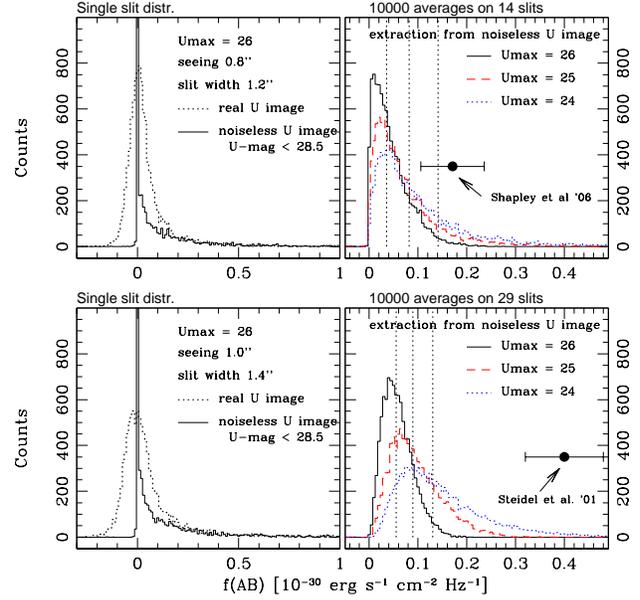}}
\caption{{\bf Upper panels:} Simulations on the LyC contamination in the case of S06.
In the left side the flux distribution of the 10000 rectangular apetures randomly positioned
over the observed U-band image (dotted line) and the noiseless image (with number counts down to $\sim$28.5, 
solid line) are shown. As expected the noiseless image
is peaked at the null flux value (corresponding to rectangular apertures not intercepting sources). In both cases the asymmetry 
towards brighter fluxes is the signature of pollution by interlopers (the case with $Umax$=26 is shown).
In the rigth panel the distributions of the averages over 14 rectangular apertures drawn from the noiseless
image are shown, in the case of Umax=24, 25 and 26 (dotted, dashed and solid lines, respectively). The vertical dotted
lines from left to right mark the median, the 1 and 2 $\sigma$ confidence levels in the case Umax=26. 
The estimation reported by S06 (average of 14 spectra) is also shown with a filled circle.
{\bf Lower panels:} The same as upper panels, but the simulations have been computed for the case of
Steidel et al. (2001).}
\label{MagInteg}
\end{figure}

In the case of the stacking of LyC non-detections the foreground
contamination may play a role in the result. The stacking process increases the signal to noise ratio
of the possible true emission of LyC, but at the same time a bias by foreground faint sources (not
detected singularly) may arise. We calculate the expected average integrated contribution of the
foreground blue sources in different rectangular apertures performing Monte-Carlo simulations.
In order to avoid possible contribution
of the error fluctuation of the background, a noiseless image has been generated. It has been obtained by running the
SExtractor algorithm (Bertin \& Arouts 1996) allowing the detection down to U-mag$\simeq$28.5 and 
requiring the ``OBJECT'' output image (CHECKIMAGE$\_$TYPE=OBJECT),
that contains the detected sources separated by pixels with null values.
10000 rectangular apertures have been randomly distributed over the noiseless and the original ultradeep VIMOS U-band images.
On one side, the rectangle width ($W$) mimic the slit width 
and on the other side, the rectangle length ($L$, spatial direction) represents the inability to 
discriminate foreground sources closer than the seeing ($L$=2$\times$$seeing$). S06 and Steidel et al. (2001)
reported a typical seeing of $\gtrsim$ 0.8\see~and used a slit width of 1.2\see~and 1.4\see, respectively.
Considering these values the simulations have been performed adopting rectangular sizes ($W$$\times$$L$) of
1.2\see$\times$1.6\see~and 1.4\see$\times$1.6\see.
Since in the real case relatively bright neighbours are easily recognised as contaminants,
we have excluded from the possible interception sources brighter than a given U magnitude
$Umax$ (with $Umax$=24, 25 and 26). In this respect, it is important to check which effect has this cut
in the optical magnitude domain, i.e. how many relatively bright sources in the optical are faint in the U
band, and therefore recognisable as potential contaminants in the reference band (e.g. R). The mean R magnitude
of the LBGs samples of S06 and Steidel et al. (2001) are 23.92$\pm$0.32 and 24.33$\pm$0.38, respectively. 
Using the deep VLT/VIMOS R-band catalog of Nonino et al. (2009) that reaches R-mag$\simeq$29 at 1$\sigma$ within 
the 1\see~radius aperture, it turns out that
with U-mag$>$24 (i.e. $Umax$=24) the fraction of galaxies with R-mag $<$ 24 (25) is 7\%(21\%). In the case
of $Umax$=25 and 26 the fractions are 3\%(13\%) and 1\%(5\%) with R-mag $<$ 24 (25), respectively.
Therefore, in the following we refer to the last two cases, and in particular to the case with $Umax$=26
when calculating the quantities related to the LyC contamination.
 
The panels in the left-side of Figure~\ref{MagInteg} show the flux (AB) distributions of the 10000 random 
positioned rectangular apertures. The median and the dispersion of these distributions quantify the contamination to the LyC estimation
for the single observation, the dotted line shows the distribution derived from the original
U-band image and the solid line is that of the noiseless image.
The difference between the two is clearly visible around the zero flux value, where in the case of the noiseless image there is a
sharp peak, while in the real image case the peaked is around the zero value, but with a dispersion
that is the contribution of fainter sources (U-mag$>$28.5) and background noise fluctuation (see negative side of the distribution). 
More importantly, in both cases the asymmetry towards 
brighter fluxes is the signature of pollution by intercepted sources (modulated by the $Umax$=26 in the case shown).
In the following we adopt the noiseless image for the estimation of the LyC contamination.
In order to simulate the effect in the average process, 10000 averages adopting the observational conditions of S06
(14 sources, W=1.2\see~and seeing 0.8\see) and Steidel et al. 2001 (29 sources, W=1.4\see~and seeing 1.0\see) have been
performed. The right side of Figure~\ref{MagInteg} shows these distributions drawn from the noiseless image and adopting
Umax=26, 25, 24 (solid, dashed and dotted lines, respectively). The medians and the one and two sigmas percentiles are reported
in Table~\ref{MEDIANS}. 
Adopting the number counts down to 28.5, in the Steidel et al. (2001) configuration ($W$$\times$$L$ = 1.4\see$\times$2.0\see) 
and considering those realizations with $Umax$=26, the median, the 1$\sigma$ and 2$\sigma$ lower bounds of the distribution of 
the averages are 29.56, 29.03 and 28.62, respectively (they are [29.07, 28.48, 28.03] adopting $Umax$=25).
Similarly, in the S06 conditions ($W$$\times$$L$ = 1.2\see$\times$1.6\see) and considering those 
realizations with $Umax$=26, the median, the 1$\sigma$ and 2$\sigma$ lower bounds are
30.02, 29.13 and 28.54, respectively (they are [29.62, 28.72, 27.92] adopting $Umax$=25, see Figure~\ref{MagInteg}, right panels and
Table~\ref{MEDIANS} for a summary). 

It is interesting to explore how these distributions change if the number counts are extrapolated down to
fainter flux limits (Umag=30.5). 
This has been done by fitting a straight line to the observed counts between U-mag 25 and 28 with a slope 
of 0.2 in the [$log(N)$ vs. U-mag] plane (Table~\ref{U_COUNTS} shows the extrapolated values). 
From the real U-band image (``OBJECT'' output image provided by SExtractor) 30 templates sources 
with U-mag$\simeq$27 have been extracted. Starting from this
sample the magnitude range 28.5$<$U-mag$<$30.5 has been populated by inserting randomly 
the galaxies appropriately dimmed according with the expected (extrapolated) number counts.

The resulting medians, 1 and 2 $\sigma$ of the distributions down to these new magnitude limits are reported in Table~\ref{MEDIANS}.
Increasing the U-mag limit for the number counts, it is apparent that the median values change more significantly 
than the dispersions ($\sigma$) and it is due to the fact that the probability to intercept (faint) sources significantly increases
(the probability to intercept a contaminating source is $\gtrsim$20\%). Conversely, the variation of the 
scatter ($\sigma$) is lower, since it is mainly sensitive to the brighter contaminants, as expected when an average is computed.
It is also worth noting that passing from U-mag 28.5, 29.5 and 30.5, the relative variations of the medians and the dispersions decrease, 
suggesting that a further extrapolation (e.g. down to U-mag 31.5) do not change significantly the present results.

\begin{table}
\centering 
\caption{Medians and 1, 2 $\sigma$ estimations of the integrated flux due to foreground sources for the samples of S06 (14 galaxies)
and Steidel et al. 2001 (29 galaxies). 
The values have been derived from the noiseless images with number counts down to 28.5, 29.5 and 30.5. 
The latter has been built by extrapolating the observed number counts (see text for details).}
\label{MEDIANS} 
\begin{tabular}{lcccccc} 
\hline
\hline   
\multicolumn{1}{c}{Noiseless}&\multicolumn{3}{c}{N=14} & \multicolumn{3}{c}{N=29}      \\
\multicolumn{1}{c}{Image}    & Median& 1$\sigma$ & 2$\sigma$ & Median&1$\sigma$& 2$\sigma$\\
\hline
Umax=25                &       &           &           &       &         &          \\
\hline
  U$\leq$28.5          & 29.62 &28.72      &27.92      & 29.07 & 28.48   & 28.03    \\
  U$\leq$29.5          & 29.25 &28.54      &27.83      & 28.74 & 28.29   & 27.90    \\
  U$\leq$30.5          & 29.13 &28.49      &27.80      & 28.63 & 28.21   & 27.85     \\
\hline
\hline
Umax=26                &       &           &           &       &         &          \\
\hline
 U$\leq$28.5           & 30.02 &29.13      &28.54      & 29.56 & 29.03   & 28.62    \\
 U$\leq$29.5           & 29.53 &28.89      &28.40      & 29.08 & 28.71   & 28.39    \\
 U$\leq$30.5           & 29.38 &28.81      &28.35      & 28.95 & 28.61   & 28.31     \\
\hline
\hline
\end{tabular}
\end{table}

It is now possible to compare the observed residual flux reported in literature 
with the present simulations. Steidel et al. (2001) measured a residual flux in their composite 
spectrum of 29 LBGs of $m_{AB}$$\simeq$27.4 (AB) ($f900$(AB)=4.02$\times$$10^{-31}erg~sec^{-1}cm^{-2}Hz^{-1}$) 
and no direct detection was identified.
They reported an observed ratio $(f1500/f900)_{OBS}$ of 17.7$\pm$3.8 that corresponds to a contrast 
of 3.1 magnitudes in the AB scale. S06 using a slit width of 1.2\see~and the average over 
the 14 sources measured 28.33 (AB) also shown in the Figure~\ref{MagInteg}
(adopting their observed ratio $f1500/f900$=58 and the average magnitude of
the sample $R$=23.92). 
Considering the simulation with the noiseless image and number counts down to U-mag=28.5,
it turns out that the LyC estimations by Steidel et al. (2001) and S06 
are significant at $>$2$\sigma$ level from the expected median foreground contamination.
Adopting the resulting median flux values derived from the simulations it is now possible to correct 
the LyC value observed by S06. Assuming the case of the noiseless image with number counts down to 30.5 
and Umax = 25 and 26, 
it turns out that the corrected residual $f900$ flux would be $\sim$ 1.9 (29.04 AB) 
and 1.6 (28.85 AB) times smaller than the value observed (28.33 AB), respectively.
In the case of Steidel et al. (2001) the average residual $f900$ flux (magnitude 27.4) is brighter than S06, 
so the correction is less important. The corrected residual $f900$ flux would be $\sim$ 1.5 (27.82 AB) 
and 1.3 (27.70 AB) times smaller, adopting the medians calculated from the noiseless images 
down to U-mag 30.5 with Umax = 25 and 26, respectively.
This would also decrease the average relative escape fraction
($<f_{esc,rel}>$) by the same factors. \footnote{See Steidel et al. (2001) or S06 Eq.~1 for the definition.}

A way to further investigate the contamination occurrence due to foreground blue sources on $z\sim$ 3-4 galaxies 
is to use high-resolution, deep and multicolour images as those obtained from the GOODS project. 
This is the argument of the next section.

\section{U-band detected LBGs at z$>$3.48 in the GOODS-S}
\label{direct_GDS}
An observational test can be performed directly on the GOODS-S area by
cross-correlating the VLT/VIMOS U-band catalog with the list of available B-band dropout galaxies
observed during the VLT/FORS2 campaign with known spectroscopic redshift (Vanzella et al. (2008, 2009a))
and for which the Lyman limit is redder than the filter transmission.
The astrometric solution between the U and the ACS catalogs gives an r.m.s. in RA and DEC
lower than 0.1\see~on average (Nonino et al. 2009).
Adopting a radius of 0.5\see, 1.0\see~and 1.3\see~in the matching, it turns out that 1, 4 and 11 galaxies 
out of 36 with redshift between 3.47-4.3 have an U-band detection with U-mag in the range 26.1-29.4.\footnote{At redshift larger 
than 3.4 the 912\AA\ limit is redder than the VIMOS U-band response (Nonino et al. 2009). 
It is also known that one of the four quadrants of the VIMOS U-band image
shows a read leak at $\lambda$$\sim$4850\AA, however dedicated simulations show that it has a negligible 
effect ($\simeq$ 0.1\%) (see Nonino et al. (2009), appendix C). 
Moreover the quadrant is not affecting the cases we are discussing here, especially that on the HUDF.
In the following we consider the cases with U-mag $\lesssim$ 28.5, for which the detection is more 
reliable ($\gtrsim$3$\sigma$). Given the sample of LBGs considered ($N$=36) and adopting the number counts 
down to U$\lesssim$28.5,} the maximum of the probability distribution peaks at 1 and 4 galaxies contaminated
out of 36. Therefore the numbers we obtain matching the U-band and the ACS catalog are fully
consistent with the random occurrence of foreground superpositions (see Table~\ref{OBS_GOODS}).
\begin{figure} \centering
  \resizebox{\hsize}{!}{\includegraphics{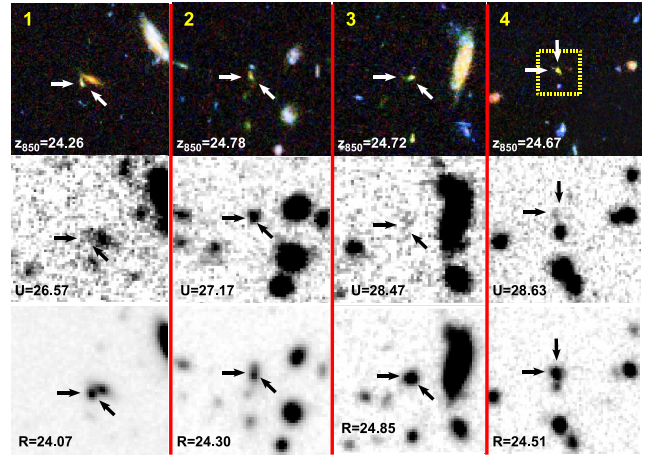}}
\caption{BVi composite ACS colour images ($top$), the VLT/VIMOS U band images ($bottom$) and
the VLT/VIMOS R band images ($bottom$) of the four cases 
(out of 36 LBGs) that satisfy the matching between
the U-band and the ACS v2.0 catalogs, adopting a radius of 1.0\see~(the cutouts 
of 10\see~size show several blue compact sources in the blank field). 
From left to right the U magnitude increases and the arrows indicate the
position of the LBGs. The upper-right case is shown with more detail 
in Figure~\ref{escape} (dotted square).}
\label{4_cases}
\end{figure}

In the following we consider the matching within 1\see~(since typically
ground-based imaging is performed with seeing better than 1\see).
In this case the
four LBGs that lie within 1.0\see~are:
(1) GDS~J033225.16-274852.6 (U=26.57), (2) GDS~J033220.97-275022.3 (U=27.17), (3) GDS~J033226.28-275245.7 (U=28.47)
and (4) GDS~J033236.83-274558.0 (U=28.63) (see Figure~\ref{4_cases}).
The magnitudes of the involved U-band detected sources range within 26.57$<$U$<$28.63, corresponding to a percentage 
of the observed continuum flux of the LBGs in the \wi\ band ($\sim$1600-1700\AA~rest-frame) of 2-10\%,
a quantity that resemble the observed LyC values often reported in literature.
Apart from the brighter case (1) in which the contamination comes from
the relatively bright galaxy close to the LBG (GDS~J033225.09-274852.6, \wz\ = 22.61, \wb\ = 26.0), 
in the other three cases the depth of the GOODS (and HUDF) area allow to visually appreciate
the presence of at least one compact blue source (not distinguishable in the ground-based $R$ band) shifted 
to respect the targeted LBG and within 1\see~distance. In particular in the 
case (4), the fainter in the U-band among the four discussed, the spatial offset is $<$ 0.5\see, and 
sits in the HUDF. 
In the next section we report on this (extreme) system as an example of LyC
contamination.
\begin{table}
\centering 
\caption{Probabilities \% P($K$) and P($\ge$$K$)) that ``$K$'' and ``at least $K$'' galaxies out of 36 ($N$) 
in the GOODS-S field used to check the occurrence of foreground contamination within 
0.5\see~and 1.0\see~are reported (assuming U-band number counts down magnitude 28.5). 
The number of cases observed (1/36 with 0.5\see~and 4/36 with 1.0\see~radius) are highlighted in bold face.}
\label{OBS_GOODS} 
\begin{tabular}{c|cc|cc} 
\hline
\hline   
                    &     P($\ge$$K$)    &  P($\ge$$K$)  &     P($K$)    &  P($K$)     \\
      K   & 0.5\see~radius     &  1.0\see~radius & 0.5\see~radius     &  1.0\see~radius\\  
\hline
 0        & 100.0           & 100.0   &   31.6       & 0.8  \\ 
 1        & {\bf 68.4}      &  99.2   &   {\bf 37.0} &  4.1  \\  
 2        & 31.4            &  95.2   &   21.1       & 10.2  \\
 3        & 10.4            &  84.9   &   7.8       & 16.8  \\
 4        & 2.6             &  {\bf 68.2}& 2.1       & {\bf 19.9} \\
 5        & 0.5             &  48.2   &  0.4         & 18.4  \\
\hline
\hline
\end{tabular}
\end{table}

\subsection{An exemplary faint foreground contamination at z=3.797 in the HUDF}
\label{HUDF_example}
\begin{figure} \centering
  \resizebox{\hsize}{!}{\includegraphics{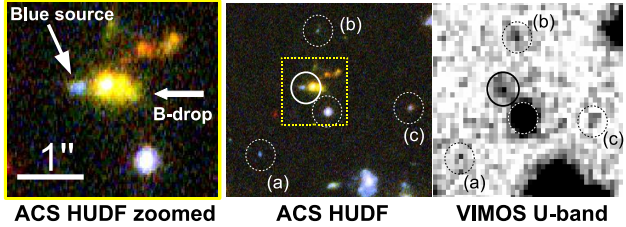}}
\caption{{\bf Left:} Detail of ACS HUDF colour image of the B-band dropout
galaxy (case 4 in Figure~\ref{4_cases}) at redshift 3.797. 
A blue compact source at $\sim$0.4\see~from the center of the LBG is indicated, 
superimposed to the high redshift galaxy. {\bf Middle:} 10\see~side field around the LBG.
The yellow dotted squared is shown in the left panel. Other three compact U-detected sources 
are marked, a), b) and c). {\bf Right:} The VLT/VIMOS U-band image of the middle panel (Nonino et al. 2009). 
The detection at the position of the high redshift galaxy is clear (5$\sigma$, solid circle), in which the presence of 
a non-resolved (in the U-band) blue compact source is evident in the HUDF image (left side).}
\label{escape}
\end{figure}

An example of $false$ detection of escape radiation of one of the B-band dropout
galaxies discussed above 
is shown in the right panel of Figure~\ref{escape}. The clear detection in the U 
band is shown (U magnitude 28.63$\pm$0.2) at the position of the galaxy 
GDS~J033236.83-274558.0 at redshift 3.797.  In the left panel of the same figure the
colour image extracted from the HUDF is shown. It is clearly visible a
blue compact source superimposed on the $z$=3.797 galaxy slightly offset 
by $\lesssim$0.4\see~(this blue source is visible also in the ACS/B435 and V606 bands).
We note that the LBG and the blue source have been detected as a single galaxy in the
HUDF catalog by Coe et al. (2006). If the blue spot is not a contaminant, then it would be 
an ultradeep, high resolution ``morphologically detected'' LyC case at high redshift.
It is not clear what would be the morphological appearance of galaxies showing LyC escape photons at 
$\lambda<$ 912\AA, however the system can be easily explained by the presence of a foreground source rather
than a shifted ``hole'' where Lyman photons can escape.
Therefore we favour the simpler interpretation of a superposition case.
It is worth noting that this system would appear as a single source if observed with lower spatial resolution, as in the
$R$ band or deep U image (seeing 0.8\see, where presumably the blue source is entirely contributing
to the U-band detection).
A deep spectroscopic or narrow band observation of this system with a
seeing larger than 0.5\see~would
detect a signal below 912\AA\ erroneously ascribed to the source at higher redshift.
In particular, given the U mag of 28.63 and \wi\ = 24.67 the polluting source mimic
an observed $(f1500/f900)_{OBS}\sim$ 38. This value is at least three times 
larger than the current LyC detections (S06 and I09), and comparable to the reported estimates 
of stacked spectra, i.e. 17.7 and 58 reported by Steidel et al. (2001) and S06, respectively.

\section{Concluding remarks}
\label{conclusion}
Depending on the seeing conditions and/or the aperture sizes adopted to carry out the photometry/spectroscopy 
the contamination probability by foreground blue sources on the estimation of the ionising radiation may not be negligible.
In particular in the cases reported by S06 and I09 it turns out that the
probability of foreground contamination for at least half of the direct LyC detections 
is $\sim$ 30\% and 50\%, respectively. 
Among the direct LyC detections in I09 there are cases of spatially offset emission
(in the case of LBGs) and cases of extremely high escape fraction (even larger than unity). 
We argue that, other than a possible complex physical explanations (e.g. Inoue 2009),
these suspicious systems should be better investigated with higher resolution and multi-band imaging,
since they are statistically compatible with superposition effects.

Also in the case of statistical LyC detection the foreground 
contamination plays a role in the global result. A stacking process increases the signal to noise 
of the possible true LyC emission, but at the same time a bias by foreground faint sources (not
detected individually) may arise. The issue has been addressed by computing Monte-Carlo simulations from which,
taking advantage of the ultradeep VLT/VIMOS U-band number counts (Nonino et al. 2009), we calculated the 
median contribution of foreground sources to the residual flux measure below the Lyman limit. 
From these calculations it turns out that the current estimations of LyC emission from stacked 
spectra at redshift $\sim$3 are beyond 2$\sigma$ from the expected median foreground contribution.

{\it The simple foreground contamination may not entirely explain the cases of direct
escape fraction observed or the limits derived from the stacking, nevertheless the calculations 
and the examples here reported demonstrate the need for caution in LyC measurements
at high redshift}. This systematic error has to be considered when attempting to measure the 
LyC escape fraction and its evolution with redshift and the contribution of galaxies to the UV background.
For example, from simulations it turns out that the corrected observed $f900$ flux reported by S06 
would be $\sim$ 1.6--1.9 times smaller than the reported value.
Differently, the correction for the Steidel et al. (2001) estimation is less effective (a factor $\sim$ 1.3), 
since their residual flux is $\sim$ 1 magnitude brighter than S06. Consequently the relative escape fractions 
are dimmed by the same factors.
It is worth noting that in stacked images of tens and hundreds galaxies at redshift $\sim$ 1
there is currently no direct detection (Cowie et al. (2009), Siana et al. (2007, 2010), Malkan et al. (2003)).
On the one hand, it may reflect indirectly the redshift distribution
of the (faint U-mag$>$26) blue compact sources that populate the ultradeep U-band image. 
Since not evident contaminations have been currently measured at
redshift $\sim$ 1, the bulk of the (faint) blue compact sources would be at redshift $>$ 1.
On the other hand, being these objects U-band detected, their typical redshift is most probably smaller 
than 3 (i.e. they are not U-band $dropouts$). For the brighter part of the distribution (U-mag$<$26.5), 
a further constraint on the redshift comes from the MUSIC photometric redshift catalog (Grazian et al. 2006),
from which they span the interval 0$<$z$<$3.

Conversely, in the case of LyC observations at z$>$3 the contribution of such blue
galaxies as contaminants largely increases.
In general, the fraction of galaxies contaminated
ranges between $\sim$3 and $\sim$15\%, varying from seeing 0.5\see~to 1.0\see, respectively,
and assuming number counts down to U-mag$\sim$28.5. On the one hand simulations predict
an evolution (increase) of the escape fraction with increasing redshift (e.g. Razoumov \& Sommer-Larsen 2007),
on the other hand also the contamination by foreground sources it is expected to increase, 
with a more significant effect for fainter LBGs ($R>24$).
Therefore, if the contamination is not taken into account properly, the observational constraints 
on the evolution of the escape fraction may be biased, particularly at faint limits where
the population of blue compact sources increases in number density.
In order to produce a robust LyC measurements at z$\gtrsim$3 high spatial resolution, multi-band and ultradeep 
imaging are necessary to exclude the spurious cases.

\section*{Acknowledgments}
 We would like to thank the referee (I. Iwata) for very constructive comments 
 and suggestions. EV would like to thank M. Giavalisco for useful comments and discussions
 about this work.
 We are grateful to the ESO staff in Paranal and Garching who greatly helped
 in the development of this programme. 
 EV acknowledge financial contribution from contract ASI/COFIN I/016/07/0 and 
 PRIN INAF 2007 ``A Deep VLT and LBT view of the Early Universe''.

\bsp
\label{lastpage}
\end{document}